\documentclass[a4paper,prc,nofootinbib,superscriptaddress,twocolumn,a4paper]{revtex4}

\usepackage{graphicx}
\usepackage{color}
\usepackage{amssymb}

\begin{document}
\title{Neutron skin of $^{27}$Al with Skyrme and Korea-IBS-Daegu-SKKU density functionals}

\author{Hana Gil}
\affiliation{Center for Extreme Nuclear Matters, Korea University, Seoul 02841, Korea}
\author{Chang Ho Hyun}
\email{hch@daegu.ac.kr}
\affiliation{Department of Physics Education, Daegu University, Gyeongsan 38453, Korea}
\author{K. S. Kim}
\email{kyungsik@kau.ac.kr}
\affiliation{School of Liberal Arts and Science, Korea Aerospace University, Goyang 10540, Korea}

\date{\today}

\begin{abstract}

Recent measurement of the parity-violating (PV) asymmetry in the elastic electron scattering on $^{27}$Al target
evokes the interest in the distribution of the neutron in the nucleus.
In this work, we calculate the neutron skin thickness ($R_{np}$) of $^{27}$Al with nonrelativistic nuclear structure models.
We focus on the role of the effective mass, symmetry energy and pairing force.
Models are selected to have effective masses in the range $(0.58-1.05)M$ where $M$ is the nucleon mass in free space,
and stiffness of the symmetry energy is varied by choosing the slope of the symmetry energy in the range 9.4 -- 100.5 MeV.
Effect of pairing force is investigated by calculating $R_{np}$ with and without pairing, and using two different forms of the pairing force.
With nine models, we obtain $R_{np} = 0.001 - 0.014$ fm.
The result is independent of the effective mass, symmetry energy, and the form of pairing force.
However, $R_{np}$ is negative when the pairing force is switched off,
so the pairing force plays an essential role to make $R_{np}$ positive and constrained in a narrow range.
We also calculate the PV asymmetry ($A_{\rm pv}$) in the elastic electron-$^{27}$Al scattering
in the Born approximation at the kinematics of the Qweak experiment.
We obtain a very narrow-ranged result $A_{\rm pv} = $ (2.07 -- 2.09) $\times 10^{-6}$.
The result is consistent with the experiment and insensitive to the effective mass,
symmetry energy and pairing force.

\end{abstract}
\maketitle




{\it Introduction.}
Measurement of parity violation in the elastic electron scattering on nuclei is expected to
shed light on the distribution of neutrons within an atomic nucleus.
For the neutron-rich nuclei neutron skin thickness ($R_{np}$) is known to be strongly correlated to the density dependence of the symmetry energy.
With precise measurement of the parity-violating (PV) asymmetry ($A_{\rm pv}$), the neutron skin
thickness of the target nuclei can be determined, and it can constrain the density dependence of the symmetry energy.
It is shown that the slope parameter $L$ of the symmetry energy
is most closely correlated to the neutron skin thickness of $^{208}$Pb \cite{ksym}.

Recently, measurement of $A_{\rm pv}$ with $^{27}$Al has been performed by the Qweak Collaboration \cite{prl2022}.
Major goal of the Collaboration is to determine proton's weak charge from the elastic $\vec{e} p$ scattering.
Largest background of the experiment comes from the aluminum alloy cell that contains the hydrogen target.
To extract an accurate value of proton's weak charge, background should be understood precisely.
Measurement of $A_{\rm pv}(^{27}{\rm Al})$ was guided by a theoretical work \cite{prc2014},
in which $A_{\rm pv}(^{27}{\rm Al})$ is calculated with the FSUGold relativistic mean field (RMF) model.
The experimental result is obtained in good agreement with the predictions in \cite{prc2014}.
From the measured value $A_{\rm pv} = (2.16 \pm 0.11 \pm 0.16) \times 10^{-6}$,
the neutron skin thickness of $^{27}$Al is obtained as $R_{np} = -0.04 \pm 0.12$ fm.
Since there is only one more neutron than the proton in $^{27}$Al, neutron skin will not be thick, and the result turns out to be consistent with the guess.
However, the range of $R_{np}$ determined from $A_{\rm pv}(^{27}{\rm Al})$ is significantly uncertain containing both negative and positive values.
Moreover, in the theoretical work \cite{prc2014}, only one RMF model is used in the calculation of $A_{\rm pv}$.
For an improved constraint on the neutron skin thickness of $^{27}$Al,
a systematic investigation is required.

In recent works, we applied the KIDS (Korea-IBS-Daegu-SKKU) density functional \cite{kidsnm} to the quasielastic electron and neutrino scattering off nuclei, and showed that the model reproduces the experimental data at high accuracy \cite{kids-scatt1, kids-scatt2, kids-scatt3}.
In the present work, we calculate $R_{np}$ of $^{27}$Al and $A_{\rm pv}$ in the elastic electron-$^{27}$Al scattering with nonrelativistic energy density functionals (EDFs).
Nuclear models consist of seven Skyrme forces and two KIDS models.
Nuclear properties in the vicinity of surface could be sensitive to the momentum-dependent terms in the Skyrme force.
Contribution of the momentum-dependent terms appears directly in the effective mass of the nucleon,
so the role of these terms to $R_{np}$ and $A_{\rm pv}$ is examined by choosing models with effective masses in a wide range at the saturation density.
Since $^{27}$Al is an open-shell and odd-number nucleus, the effect of pairing can play a critical role in the properties of the nucleus.
Dependence on the pairing force is examined by adopting two different forms of the pairing force:
one is the constant gap form, and the other is the constant force form.
In addition, by switching the contribution of the pairing on and off, the effect of pairing force is thoroughly explored.

In the calculation of $A_{\rm pv}$, following the prescription in \cite{prc2014}, we assume the distribution of proton and neutron spherically symmetric, and calculate the multipole form factors of the proton by using the hole state in the $d_{5/2}$ level.
With the nine EDF models, $A_{\rm pv}$ is calculated at the laboratory angle $\left< \theta_{\rm lab} \right> = 7.61^\circ \pm 0.02^\circ$ and incident electron energy $\left< E_{\rm lab} \right> = 1.157$ GeV.


\begin{table*}
\begin{center}
\begin{tabular}{cccccccccc} \hline
 & SkI3 & SkMP & SLy4 & KDE0v1 & SkM* & LNS & KIDS0-m*97 & KIDS0 & MSk7\\ \hline
$m^*_s/M$ & 0.58 & 0.65 & 0.69 & 0.74 & 0.79 & 0.83 & 0.90 & 0.99 & 1.05 \\
$m^*_v/M$ & 0.80 & 0.59 & 0.80 & 0.81 & 0.65 & 0.73 & 0.70 & 0.81 & 1.05 \\
$K_0$ & 258.2 & 230.9 & 229.9 & 227.5 & 216.6 & 210.8 & 240.0 & 240.0 & 231.2 \\
$J$ & 34.8 & 29.9 & 32.0 & 34.6 & 30.0 & 33.4 & 32.8 & 32.8 & 28.0\\
$L$ & 100.5 & 70.3 & 45.9 & 54.7 & 45.8 & 61.5 & 49.1 & 49.1 & 9.4 \\
Ref. & \cite{ski3} & \cite{skmp} & \cite{sly4} & \cite{kde} & \cite{skmstar} & \cite{lns} & \cite{kids0a,kids0b,kids0c}
& \cite{kids0a,kids0b,kids0c}  & \cite{msk7} \\
\hline
\end{tabular}
\end{center}
\caption{Skyrme force models considered in the work. $m^*_s$ and $m^*_v$ denote the isoscalar and
isovector effective masses at the saturation density, respectively.
Incompressibility in symmetric nuclear matter $K_0$, and symmetry energy parameters $J$, $L$ are in units of MeV.}
\label{tab1}
\end{table*}

{\it Model and Formalism.}
Since the role of symmetry energy and momentum-dependent interaction
is a focus of the work, models having wide ranges for these quantities are selected.
In Tab.~\ref{tab1} we summarize the effective masses at the saturation density
and parameters that determine the density dependence of the nuclear matter equation of state for the nine Skyrme and KIDS models.
Models are chosen to have the isoscalar ($m^*_s$) and isovector ($m^*_v$) effective masses distributed uniformly in a wide range $(0.58-1.05)M$.
Incompressibility of the symmetric matter $K_0$ is within the range of experimental uncertainty \cite{k0}.
Density dependence of the symmetry energy is characterized by the slope $L$ in the vicinity of the saturation density.
SkI3 and MSk7 models have anomalously large and small $L$, but the other models give values of $L$ within a range consistent with neutron star observation \cite{ksym, symene}.

Pairing force is essential in explaining the odd-even staggering of the mass and charge radius.
To see the dependence on the pairing, we employ two different forms of the pairing correlation.
One is the constant gap formula
\begin{equation}
\Delta_q = \frac{11.2}{\sqrt{A}},
\label{eq:pgap}
\end{equation}
where $A$ denotes the mass number,
and the other is the constant force
\begin{equation}
G_q = \frac{29}{A}
\label{eq:pforce}
\end{equation}
in the gap equation $\Delta_q = G_q \sum _{\beta \in q} \sqrt{w_\beta (1-w_\beta)}$
where $w_\beta$ denotes the occupation probability in a given orbit.
In addition to using two different forms for the pairing correlation,
role of the pairing is investigated by calculating $R_{np}$ with and without the pairing.

Parity-violating asymmetry $A_{\rm pv}$ is defined by
\begin{equation}
A_{\rm pv} = \frac{d \sigma_+/d \Omega - d\sigma_-/d\Omega}{d \sigma_+/d \Omega + d\sigma_-/d\Omega},
\end{equation}
where the subscripts $+$ and $-$ denote the positive and negative helicity of the incoming electrons, respectively.
If both neutrons and protons are distributed equally and symmetrically within a nucleus, $A_{\rm pv}$ is simplified to
\begin{equation}
A^0_{\rm pv} = - \frac{G_F\, Q_W}{4 \sqrt{2} \pi \alpha Z} q^2,
\end{equation}
where $G_F$ is the Fermi constant, $\alpha$ is the fine structure constant, and $Z$ is the proton number of a nucleus.
Weak charge of a nucleus $Q_W$ is given by
\begin{equation}
Q_W = Q_n N + Q_p Z,
\end{equation}
where $Q_n=-1$ and $Q_p = 1- 4 \sin^2 \theta_W$ at the tree level, and $N$ is the neutron number of a nucleus.
$q^2$ is the three momentum transfer $q^2 =  4 E_i E_f \sin^2 \frac{\theta}{2} + (E_i - E_f)^2$, which reduces to
\begin{equation}
q^2 = 4 E^2_i \sin^2 \frac{\theta}{2}
\end{equation}
in the elastic scattering.


In reality, however, distribution of the proton is different from that of the neutron for various reasons.
Taking into account the difference of the neutron and proton distribution,
$A_{\rm pv}$ is given in the Born approximation as \cite{prc2014}
\begin{equation}
A_{\rm pv} = A^0_{\rm pv}\frac{C_0 C^W_0 + C_2 C^W_2 + C_4 C^W_4}{C^2_0 + C^2_2 + C^2_4}.
\end{equation}
$C_0$, $C_2$ and $C_4$ are the form factors of the proton defined by
\begin{eqnarray}
C_0(q) &=& \frac{1}{Z} \int d^3 r \rho^p_0(r) j_0(q r), \\
C_2(q) &=& \frac{1}{Z}\sqrt{\frac{10}{7}} \int d^3r \frac{1}{4\pi}|R^p_{d 5/2} (r)|^2 j_2(qr), \\
C_4(q) &=& \frac{1}{Z}\sqrt{\frac{2}{7}} \int d^3r \frac{1}{4\pi}|R^p_{d 5/2} (r)|^2 j_4(qr),
\end{eqnarray}
where $\rho^p_0(r)$ is the proton distribution in the spherical approximation,
and $R^p_{d 5/2}(r)$ is the radial function of the proton wave function in the $d_{5/2}$ state.
Weak counterparts of the form factors are
\begin{eqnarray}
C^W_0(q) &=& \frac{1}{Q_W} [Q_n N F_n(q) + Q_p Z C_0(q)], \\
C^W_2(q) &=& \frac{Q_p}{Q_W} Z C_2(q), \\
C^W_4(q) &=& \frac{Q_p}{Q_W}  C_4(q),
\end{eqnarray}
where $F_n(q) = \frac{1}{N} \int d^3 r \rho^n_0(r) e^{i \vec{q} \cdot \vec{r}}$
and $\rho^n_0(r)$ is the spherical neutron distribution.

Model dependence and validity of approximations can be checked by comparing the elastic cross section with experimental data.
Cross section of the unpolarized electrons can be approximated as
\begin{equation}
\frac{d \sigma}{d \Omega} \simeq \sigma_m F_L(q)^2,
\end{equation}
where $\sigma_m$ is the Mott cross section
\begin{equation}
\sigma_m = \frac{Z^2 \alpha^2}{4 E^2_i} \frac{\cos^2 \frac{\theta}{2}}{\sin^4 \frac{\theta}{2}},
\end{equation}
and the longitudinal form factor is
\begin{equation}
F_L(q)^2 = G_E(q)^2 F_p(q)^2.
\end{equation}
$G_E(q)$ is the electric form factor of the proton, and $F^2_p$ is obtained from
\begin{equation}
F_p(q)^2 = C_0(q)^2 + C_2(q)^2 + C_4(q)^2.
\end{equation}
Exact form of $G_E(q)$ is adopted from the quasielastic electron scattering \cite{prc2003},
where $G_E(q) = 1/(1+q^2/\Lambda^2)^2$ and $\Lambda^2 = 0.71$ (GeV/$c$)$^2$.


\begin{figure}
\begin{center}
\includegraphics[width=7.5cm]{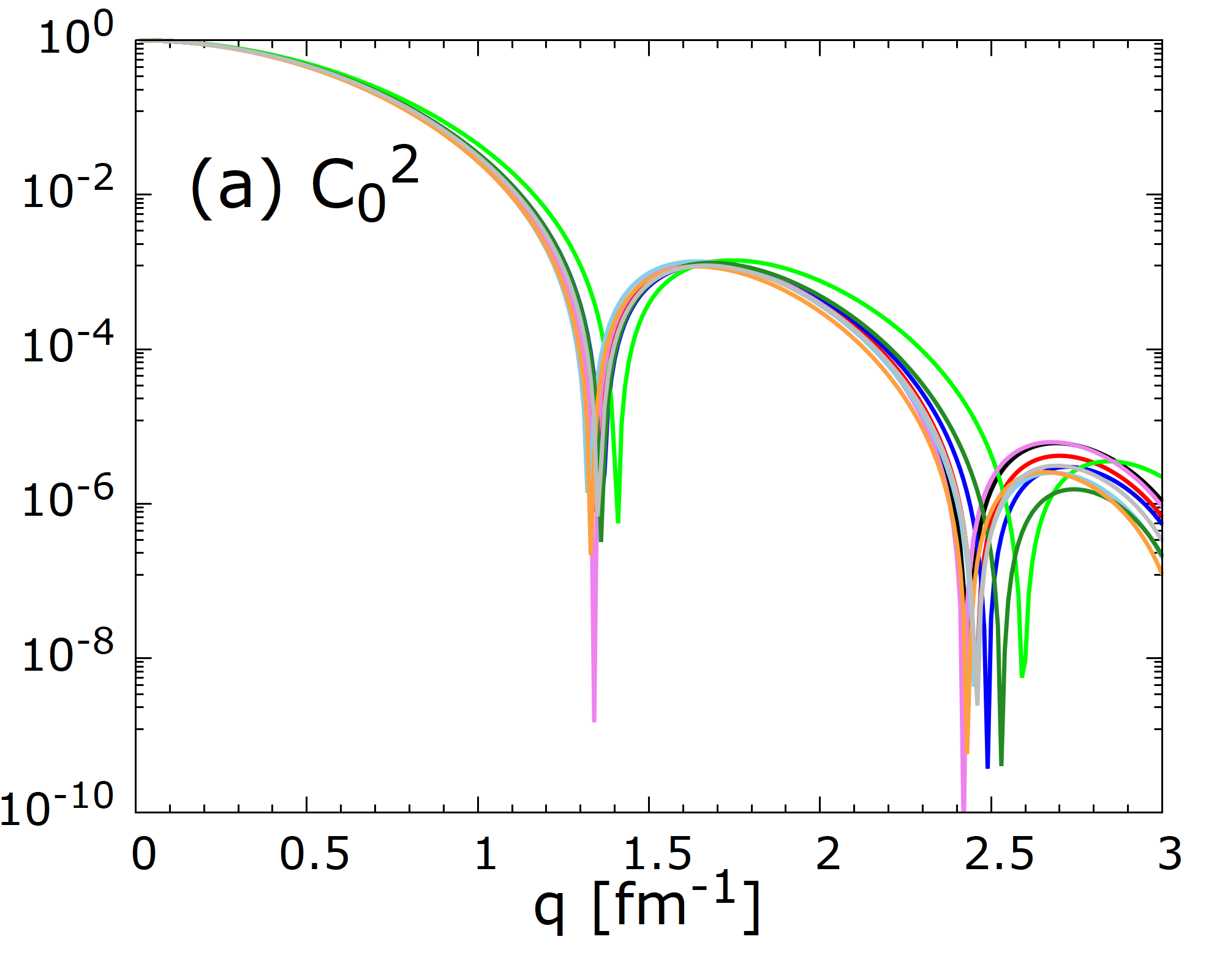}
\includegraphics[width=7.5cm]{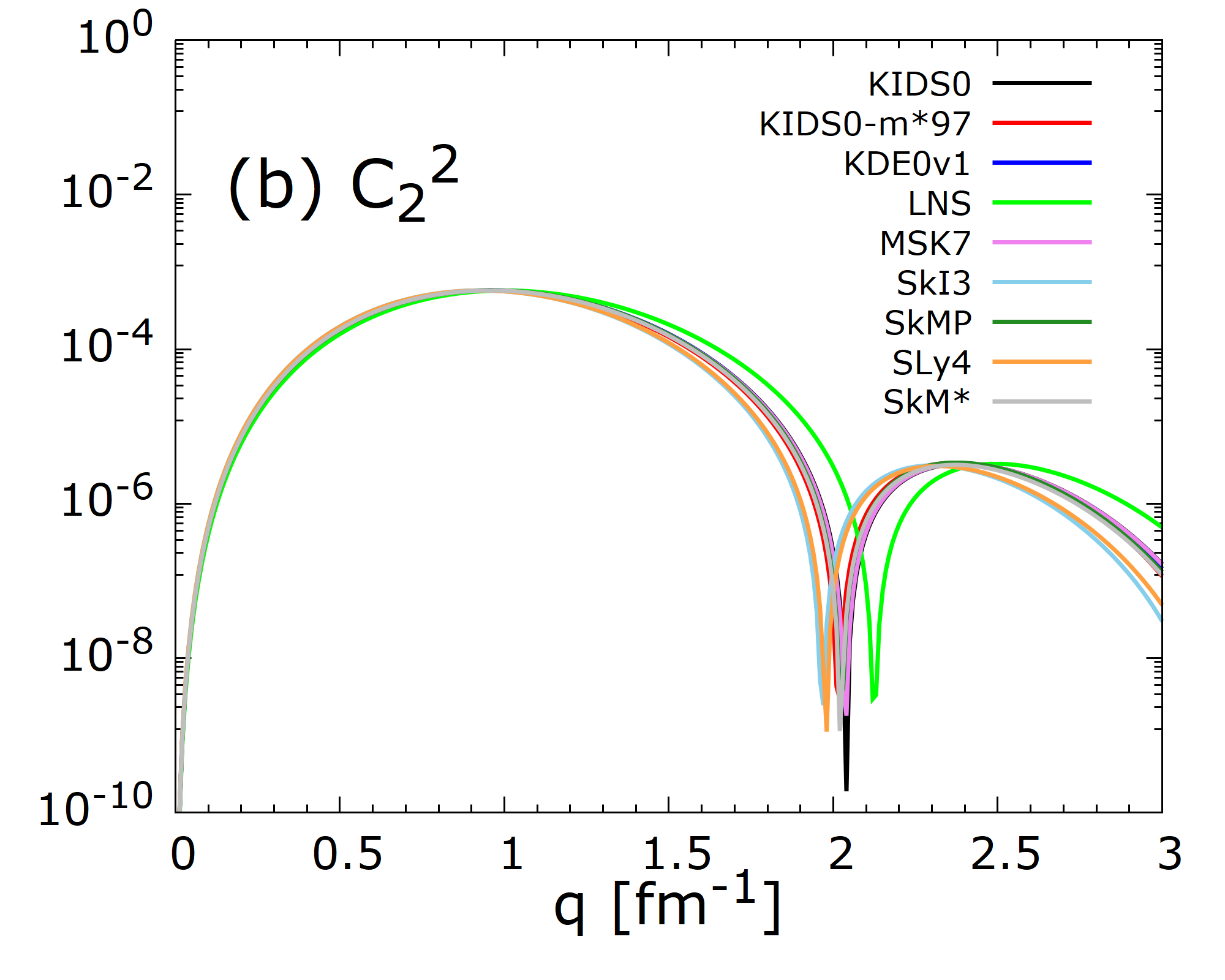}
\end{center}
\caption{Square of the form factor (a) $C^2_0$ and (b) $C^2_2$ for the nine EDF models.}
\label{fig1}
\end{figure}

{\it Result.}
Figure \ref{fig1} shows the squares of $C_0$ and $C_2$ as functions of three momentum transfer $q$ for the nine EDF models.
Order of magnitude of each term, and the dependence on $q$ are consistent with the results in \cite{prc2014, prc2017}.
At $q \lesssim 1$ fm$^{-1}$, models give very similar result,
but as $q$ increases, dependence on the model becomes evident.
In our calculation of $A_{\rm pv}$, $q \simeq 0.78$ fm$^{-1}$, so the result is expected to be weakly dependent on the model.

\begin{figure}
\begin{center}
\includegraphics[width=7.5cm]{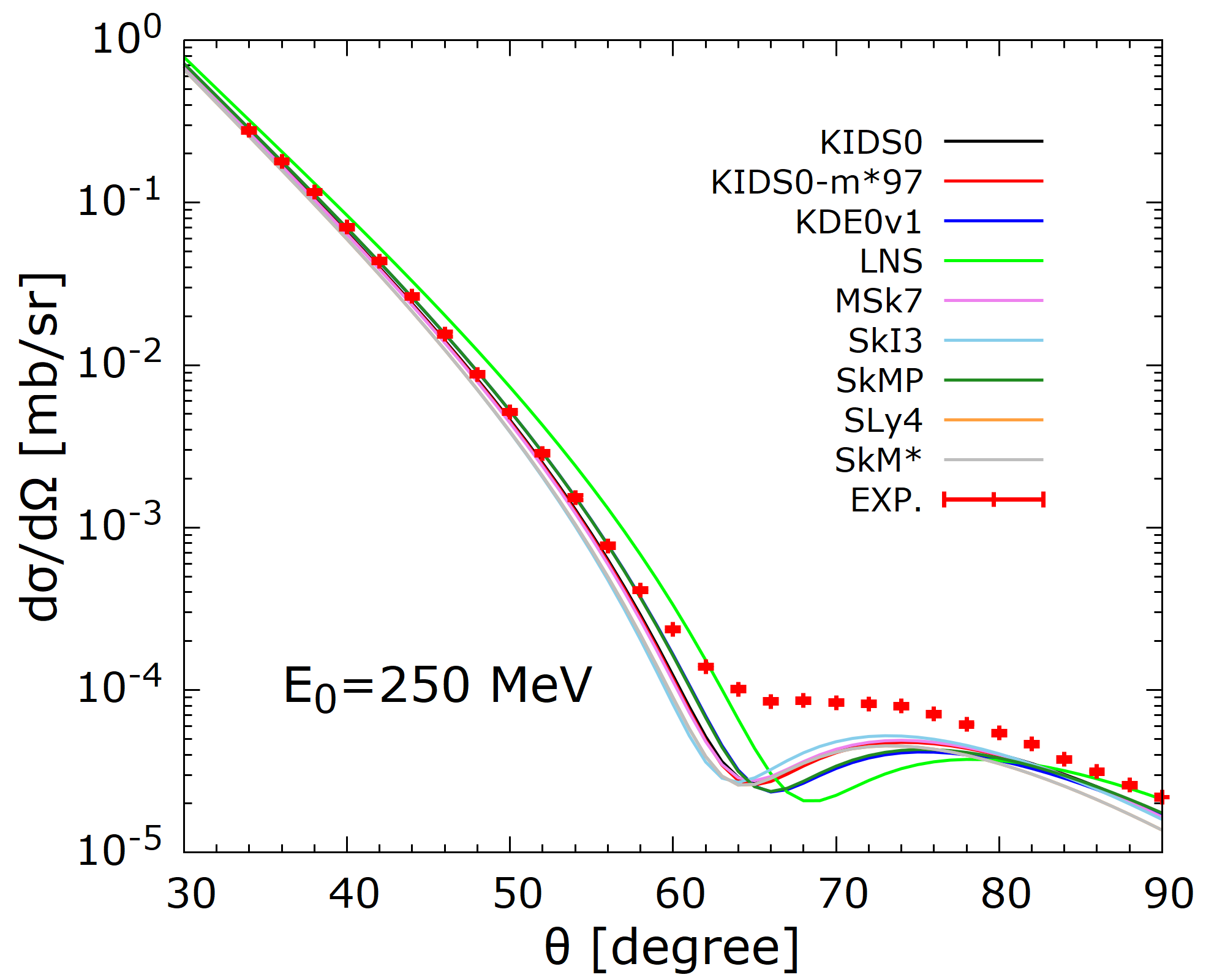}
\end{center}
\caption{Cross sections for the elastic scattering of electrons off $^{27}$Al at $E=250$ MeV.
Experimental data are taken from \cite{prc1974}.}
\label{fig2}
\end{figure}

Figure \ref{fig2} presents the cross section in the elastic electron scattering on $^{27}$Al at $E=250$ MeV as a function of the laboratory angle $\theta$.
Experimental data are available in \cite{prc1974}.
Up to 50 degrees, eight models give very similar result and agree well with data,
but the LNS model overshoots the experiment slightly.
From about middle of 50 degrees, model dependence becomes transparent.
Models are divided into four groups, and the grouping persists to about 64 degrees.
SkI3 and SkM* are located in the lowest place,
and KIDS0, KIDS0-m*97, MSk7 and SLy4 models are overlapping very closely above the SkI3 and SkM* group.
KDE0v1 and SkMP models show the best agreement to data until 58 degrees,
and LNS model is located at the highest position.
Angle that corresponds to $q=0.78$ fm$^{-1}$ at which $A_{\rm pv}$ is calculated is about 40 degrees.
At this angle, all the models agree well with data and model dependence is negligible.

\begin{table*}
\begin{center}
\begin{tabular}{ccccccccccr} \hline
                           & SkI3 & SkMP & SLy4 & KDE0v1 & SkM* & LNS & KIDS0-m*97 & KIDS0 & MSk7  & Exp. \\ \hline
 & \multicolumn{9}{c}{constant gap pairing Eq.~(\ref{eq:pgap})} & \\ \hline
$E/A$ & 7.782   & 8.651   & 7.997 & 8.564 & 8.461 & 8.562 & 8.200 & 8.273 & 8.288 & 8.332 \\
$R_n$  & 3.028   & 2.939  & 3.025 & 2.941 & 2.967 & 2.831 & 2.994 & 2.985 & 2.989 & \\
$R_p$  & 3.014   & 2.936  & 3.020 & 2.938 & 2.966 & 2.823 & 2.981 & 2.974 & 2.985 & \\
$R_c$  & 3.115   & 3.035  & 3.121 & 3.038 & 3.066 & 2.937 & 3.080 & 3.072 & 3.083 & 3.061\\
$R_{np}/10^{-2}$ & 1.32 & 0.31 & 0.43 & 0.28 & 0.10 & 0.82 & 1.30 & 1.07 & 0.37 & $-4 \pm 12$  \\
$A_{\rm pv}/10^{-6}$ & 2.068  & 2.089  & 2.086 & 2.090 & 2.092 & 2.093 & 2.076 & 2.079 & 2.091 & 1.89 -- 2.43 \\ \hline
 & \multicolumn{9}{c}{no pairing} & \\ \hline
$R_{np}/10^{-2}$ & $-3.94$ & $-3.31$ & $-3.57$ & $-3.82$ & $-3.59$ & $-3.81$ & $-3.17$ & $-3.38$ & $-3.54$ &  \\ \hline
 & \multicolumn{9}{c}{constant force pairing Eq.~(\ref{eq:pforce})} & \\ \hline
$R_{np}/10^{-2}$ & 1.38 & 0.38 & 0.51 & 0.28 & 0.14 & 0.92 & 1.33 & 1.13 & 0.42 &  \\
$A_{\rm pv}/10^{-6}$ & 2.067  & 2.087  & 2.084 & 2.089 & 2.091 & 2.092 & 2.073 & 2.076 & 2.090 &  \\ \hline
\end{tabular}
\end{center}
\caption{Model results for the properties of $^{27}$Al and parity-violating asymmetry.
Binding energy per nucleon $E/A$ is in MeV, and radii are in fm.
Experimental values of $E/A$ and $R_c$ are taken from AME2020 \cite{ame2020}
and those of $R_{np}$ and $A_{\rm pv}$ are from \cite{prl2022}.
Two rows in the bottom show the result with constant force form of the pairing.}
\label{tab2}
\end{table*}

Table \ref{tab2} displays the results for the properties of $^{27}$Al and $A_{\rm pv}$.
Table is divided into three sections: upper section corresponds to the result with constant gap formula Eq.~(\ref{eq:pgap}),
middle section is without pairing, and the lowest section is with constant force formula Eq.~(\ref{eq:pforce}).
Data of the binding energy per nucleon ($E/A$) and the charge radius ($R_c$) are taken from AME2020 \cite{ame2020},
and those of $R_{np}$ and $A_{\rm pv}$ are from \cite{prl2022}.
$R_c$ are also furnished with good agreement to the experimental data.
Deviations from the data defined by
\begin{equation}
D_{\cal O} = \left|\frac{{\cal O}_{\rm cal} - {\cal O}_{\rm exp}}{{\cal O}_{\rm exp}}\right|
\end{equation}
are less than 1 \% in six models for $R_c$, but for $E/A$ only two models KIDS0 and MSk7 satisfy the accuracy.
Neutron radius $R_n$ takes a range 2.84--3.03 fm, which is comparable to the values obtained
from the PV experiment 2.77--3.01 fm \cite{prl2022}.
Model dependence of $R_n$ is non-negligible, but with the models satisfying $D_{R_c} < 1$ \%,
$R_n$ is constrained in the range 2.94--2.99 fm, so the uncertainty due to the model reduces significantly.

For the skin thickness, we obtain the result both with and without pairing.
Without pairing the skin thickness is negative in all the models, and the magnitudes are similar.
Including the pairing force, $R_{np}$ becomes positive and determined in a range $0.001 \lesssim R_{np} \lesssim 0.014$~fm.
Our result is consistent with the result of PV experiment, and constrains $R_{np}$ in a narrower range.
Dependence on the model is, contrary to $R_n$, not affected by the accuracy of $R_c$,
so the six models that satisfy $D_{R_c} < 1$ \% have the same range of $R_{np}$ as the nine models.

Parity-violating asymmetry is obtained within a range $2.07 \times 10^{-6} \lesssim A_{\rm pv} \lesssim 2.09 \times 10^{-6}$.
There is no particular correlation with the effective mass or the symmetry energy.
The result is almost a constant, so we could not extract the range of $R_n$ from the correlation between $A_{\rm pv}$ and $R_n$ as is done in \cite{prl2022}.
Survey with nine EDF models shows that $R_{np}$ and $A_{\rm pv}$ of $^{27}$Al are determined precisely and model-independently.
They are weakly correlated to the effective mass and the symmetry energy.

With constant force pairing formula in Eq.~(\ref{eq:pforce}),
$R_{np}$ and $A_{\rm pv}$ show slight increase and decrease, respectively,
compared to the result of the constant gap.
However, the ranges of these quantities are identical in both constant gap and constant force.
To the extent we consider, $R_{np}$ and $A_{\rm pv}$ turn out to be insensitive to the form of pairing force.


{\it Summary.}
Measurement of the PV asymmetry in the elastic electron scattering on $^{27}$Al
was performed by the Qweak Collaboration, and the result has been reported recently.
The experimental result agrees well with the theoretical prediction calculated with the FSUGold RMF model.
Neutron skin thickness estimated from the measured $A_{\rm pv}$ is small enough to be consistent with $N\simeq Z$ of $^{27}$Al.
However, the value is distributed in both negative and positive regions,
and it is not clear what is the dominant source that constrains $R_{np}$ in a narrow range.
Attempting to find answers to the questions, we performed a survey in the nonrelativistic model space.
We focused on the correlation of $R_{np}$ and $A_{\rm pv}$ with the effective mass, symmetry energy and pairing force.

With the nine EDF models, we obtain $0.001 \lesssim R_{np} \lesssim 0.014$~fm.
It is remarkably independent of the effective mass, symmetry energy and the form of pairing force.
The result is consistent with the range of $R_{np}$ extracted from the PV experiment,
but in our result, negative values are excluded and upper limit is much lower than the empirical value.
We note that the radii of the neutron and the proton are non-negligibly model-dependent,
but their difference $R_{np}$ falls into a narrow range.
We found that the pairing force plays a crucial role in making $R_{np}$ positive and reside in a narrow range.
$R_{np}$ calculated without pairing force is consistently negative and the magnitude is marginally dependent on the model.
Including the pairing force, neutrons are expelled outward and $R_{np}$ becomes positive.
It is remarked in \cite{prc2014} that
{\it ``Very likely theory makes a sharp prediction for $R_n - R_p$ for $^{27}$Al that is essentially independent of the density dependence of the symmetry energy''}.
Our results confirm the guess.

More interesting is the result of $A_{\rm pv}$, which takes values in $2.068$--$2.093\times10^{-6}$.
There is no correlation with effective mass, symmetry energy and pairing force.
This behavior is different from the result of RMF models in which $A_{\rm pv}$ takes a range $2.10$--$2.22\times 10^{-6}$ \cite{prl2022}.
Accurate determination of $A_{\rm pv}$ from either theory or experiment could reduce the uncertainty arising from the nuclear background,
and make the measurement of weak charge more precise.

{\it Acknowledgments.}
This work was supported by the National Research Foundation of Korea (NRF) grant funded by the Korea government
(No. 2018R1A5A1025563 and No. 2020R1F1A1052495).

\end{document}